\documentclass[conference]{IEEEtran}
\usepackage{cite}
\usepackage{amsmath,amssymb,amsfonts}
\usepackage{algorithmic}
\usepackage{graphicx}
\usepackage{textcomp}
\usepackage{xcolor}
\usepackage{mathtools}
\usepackage[caption=false,font=footnotesize]{subfig}

\def\BibTeX{{\rm B\kern-.05em{\sc i\kern-.025em b}\kern-.08em
    T\kern-.1667em\lower.7ex\hbox{E}\kern-.125emX}}

\def\bb0{{\mathbb{0}}}

\def\bb{{\mathbf{b}}}

\def\bff{{\mathbf{f}}}

\def\bs{{\mathbf{s}}}

\def\bA{{\mathbf{A}}}
\def\bB{{\mathbf{B}}}
\def\bC{{\mathbf{C}}}

\def\bF{{\mathbf{F}}}

\def\bI{{\mathbf{I}}}

\def\bZ{{\mathbf{Z}}}

\def\bbC{{\mathbb{C}}}

\def\cI{\mathcal{I}}

\def\cR{\mathcal{R}}

\def\sf0{{\mathsf{0}}}

\def\bsf0{{\bm{\mathsf{0}}}}

\newcommand{\Exp}[1]{{\mathbb{E}\left\{#1\right\}}}                     
\newcommand{\trace}[1]{{\mathrm{tr}\left\{#1\right\}}}               
\newcommand{\norm}[1]{\left\lVert#1\right\rVert}    
\DeclareMathOperator*{\argmax}{arg\,max}            

\newcommand{\rmF}{\mathrm{F}}                       
\newcommand{\rmH}{\mathrm{H}}                       

\newcommand{\NBS}{N_{\mathrm{BS}}}                       
\newcommand{\NMS}{N_{\mathrm{MS}}}                       

\newcommand{\NBSRF}{N_{\mathrm{BS,RF}}}                  
\newcommand{\nRF}{n_{\mathrm{RF}}}                  
\newcommand{\NMSRF}{N_{\mathrm{MS,RF}}}                  

\newcommand{\Ns}{N_{\rm s}}                         
\newcommand{\ns}{n_{\rm s}}                         

\newcommand{\yBS}{\mathbf{y}_{m,n}^{\mathrm{BS}}}                       
\newcommand{\yMS}{\mathbf{y}_{m,n}^{\mathrm{MS}}}              

\newcommand{\nBS}{\mathbf{n}_{m,n}^{\mathrm{BS}}}              
\newcommand{\nMS}{\mathbf{n}_{m,n}^{\mathrm{MS}}}              

\newcommand{\Pt}{P_{\rm t}}                         
\newcommand{\varMS}{\sigma_{\rm MS}^{2}}            
\newcommand{\varBS}{\sigma_{\rm BS}^{2}}            

\newcommand{\WBS}{\mathbf{W}^{\mathrm{BS}}}              
\newcommand{\FBS}{\mathbf{F}_{m}^{\mathrm{BS}}}              
\newcommand{\FBSRF}{\mathbf{F}^{\mathrm{BS,RF}}}              
\newcommand{\FBSBB}{\mathbf{F}_{m}^{\mathrm{BS,BB}}}              
\newcommand{\WMS}{\mathbf{W}_{m}^{\mathrm{MS}}}              
\newcommand{\WMSRF}{\mathbf{W}^{\mathrm{MS,RF}}}              
\newcommand{\WMSBB}{\mathbf{W}_{m}^{\mathrm{MS,BB}}}              

\newcommand{\HBM}{\mathbf{H}_{m}}              
\newcommand{\Ht}{\mathbf{H}_{m,n}^{\mathrm{t}}}                
\newcommand{\HSI}{\mathbf{H}_{\rm SI}}              

\newcommand{\alphaBMl}{\alpha_{l}}       
\newcommand{\alphat}{\alpha_{k}^\mathrm{t}}                 

\newcommand{\aBS}{\mathbf{a}_{\rm BS}}              
\newcommand{\aMS}{\mathbf{a}_{\rm MS}}              

\newcommand{\thetaBMl}{\theta_{l}}      
\newcommand{\phiBMl}{\phi_{l}}          

\newcommand{\thetat}{\theta_{k}^\mathrm{t}}                
\newcommand{\thetar}{\theta_{\rm r}}    

\newcommand{\SINR}{\mathsf{SINR}_{\nRF}}        

\newcommand{\GT}{G_{\mathrm{T},m,\ns}}        
\newcommand{\GR}{G_{\mathrm{R},\nRF}}        

\newcommand{\tauT}{\tau_{\rm T}}        
\newcommand{\tauR}{\tau_{\rm R}}        

\IEEEoverridecommandlockouts
\IEEEaftertitletext{\vspace{-1\baselineskip}}
\begin{document}

\title{Hybrid Precoding and Combining for mmWave Full-Duplex Joint Radar and Communication Systems under Self-Interference}

\author{\IEEEauthorblockN{Murat Bayraktar\IEEEauthorrefmark{1}, Nuria Gonz\'alez-Prelcic\IEEEauthorrefmark{1}, Hao Chen\IEEEauthorrefmark{2}} \IEEEauthorblockA{\IEEEauthorrefmark{1}University of California San Diego, USA \\
\IEEEauthorrefmark{2}Samsung Research America \\
Email:\{\texttt{mbayraktar,ngprelcic}\}\texttt{@ucsd.edu}, \texttt{hao.chen1@samsung.com}}
\thanks{This material is based upon work partially supported by the National Science Foundation under grant no. 2147955 and is supported in part by funds from the federal agency and industry partners as specified in the Resilient \& Intelligent NextG Systems (RINGS) program.}
}

\maketitle

\begin{abstract}
In the context of integrated sensing and communication (ISAC), a full-duplex (FD) transceiver can operate as a monostatic radar while maintaining communication capabilities. This paper investigates the design of precoders and combiners for a joint radar and communication (JRC) system at mmWave frequencies. 
The primary goal of the  design is to guarantee certain performance in terms of some sensing and communication metrics while minimizing the self-interference (SI) caused by FD operation and taking into account the hardware limitations coming from a hybrid MIMO architecture.
Specifically, we introduce a generalized eigenvalue-based precoder design that considers the downlink user rate, the radar gain, and the SI suppression. 
Since the hybrid analog/digital architecture degrades the SI mitigation capability of the precoder, we further enhance SI suppression with the analog combiner.
Our numerical results demonstrate that the proposed architecture achieves the required radar gain and SI mitigation while incurring a small loss in downlink spectral efficiency. Additionally, the numerical experiments also show that the use of orthogonal frequency division multiplexing (OFDM) radar with the proposed beamforming architecture results in highly accurate range and velocity estimates for the detected targets.
\end{abstract}


\section{Introduction}
\label{sec:intro}

Future wireless networks will integrate sensing and communication  by exploiting a single waveform \cite{Liu2022JSAC}. 
One particular network sensing mode is based on using a full-duplex (FD) communication equipment as a monostatic radar during the downlink transmission \cite{Barneto2021WC, Liyanaarachchi2021JCS, Barneto2022TCOM, Islam2022ICC, Bayraktar2023CAMSAP}. 
Considering the large bandwidth and directional beamforming typically utilized at mmWave bands \cite{Heath2016JSTSP}, FD joint radar and communication (JRC) applications have the potential to achieve precise range and angle estimations.
Although the literature on FD communication at mmWave bands is rich \cite{ValcarceFD2019, ValcarceFD2020, RobertsTWC2021, RobertsWC2021}, the specific demands of JRC applications introduce new challenges.

The primary challenge in FD systems lies in effectively reducing self-interference (SI). In the case of mmWave systems equipped with large arrays, SI suppression only through hardware design can be costly and challenging. To address this issue, beamforming techniques are employed to provide SI mitigation in addition to the physical isolation enabled by the specific FD circuit \cite{Barneto2021WC, Liyanaarachchi2021JCS, Barneto2022TCOM, Islam2022ICC, Bayraktar2023CAMSAP, ValcarceFD2019, ValcarceFD2020, RobertsTWC2021, RobertsWC2021}. 
Assuming a hybrid MIMO architecture, it is imperative to suppress the SI before the low noise amplifiers (LNAs) and analog-to-digital converters (ADCs) of the receive side of the FD system to prevent saturation and clipping.
Consequently, SI suppression should be a key consideration for the precoder/combiner design, even though many works in the integrated sensing and communication (ISAC) literature tend to overlook the SI constraint \cite{Liu2022JSAC}.

One of the biggest challenges in precoder/combiner design for FD-JRC systems involves providing a trade-off between the sensing and communication performance and the SI mitigation capability. 
In this context, the precoders at the FD base station (BS) must be designed to maximize downlink data rate while ensuring sufficient sensing performance \cite{Liu2022JSAC, Barneto2021WC} and contributing to the SI mitigation. 
In addition, the analog combiners at the FD transceiver must be designed to facilitate radar processing and also SI supression so that target reflections can be observed.

Recent work addresses the design of hybrid precoders and combiners for FD-JRC systems \cite{Barneto2022TCOM, Islam2022ICC}. 
In \cite{Barneto2022TCOM}, the focus is on FD-JRC applications involving multiple downlink mobile stations (MSs) with only line-of-sight (LoS) paths.
The designed analog/hybrid precoders yield substantial gains in both the directions of the MSs and the radar targets, with SI suppression being accomplished solely through the use of combiners at the FD BS. 
However, the SI suppression method employed, known as null-space projection (NSP), provides a digital solution for the analog combiner, which would require an implementation based on fine grain variable attenuators in addition to phase shifters, adding an extra layer of complexity to the circuit and increased cost.
In \cite{Islam2022ICC}, a hardware-based analog cancellation architecture is adopted, with residual SI being canceled by the hybrid precoders.
This work assumes that the contributions to the downlink channel are provided by a subset of the targets, which might not necessarily hold true for JRC systems. 
Moreover, the use of DFT codebooks for the design of analog precoders and combiners limits the achievable performance.

In this paper, we consider a system that involves a downlink MS and multiple targets in the environment. 
Unlike the scenario presented in \cite{Barneto2022TCOM}, the downlink channel contains multiple paths. 
Additionally, the downlink channel paths and the angles of the targets do not necessarily coincide, which differs from the assumption connsidered in \cite{Islam2022ICC}. Finally, unlike \cite{Barneto2022TCOM}, we assume a hybrid MIMO architecture based on analog beamformers implemented only with phase shifters, which is a feasible solution at mmWave.
The proposed optimization problem seeks to maximize downlink data rate while simultaneously achieving high gain at the target angle. 
We consider constraints related to SI suppression and hardware limitations, specifically the use of an analog/hybrid architecture with phase shifters.
Initially, we design precoders through a solution to a generalized eigenvalue problem, focusing on maximizing data rates while suppressing SI.
Subsequently, we coherently combine the beams that offer both high data rates and radar gain as part of the precoder design process. 
This is followed by the use of a hybrid decomposition algorithm to obtain the analog and digital precoders.
Furthermore, we design the analog combiner at the FD BS to suppress the remaining SI while ensuring sufficient gain in the target angle. 
The simulation results show that the proposed design provides a reasonable trade-off between the sensing and communication performance with a practical architecture.

\section{System Model}
\label{sec:sysmod}

\begin{figure}
    \centering
    \includegraphics[width=0.95\linewidth]{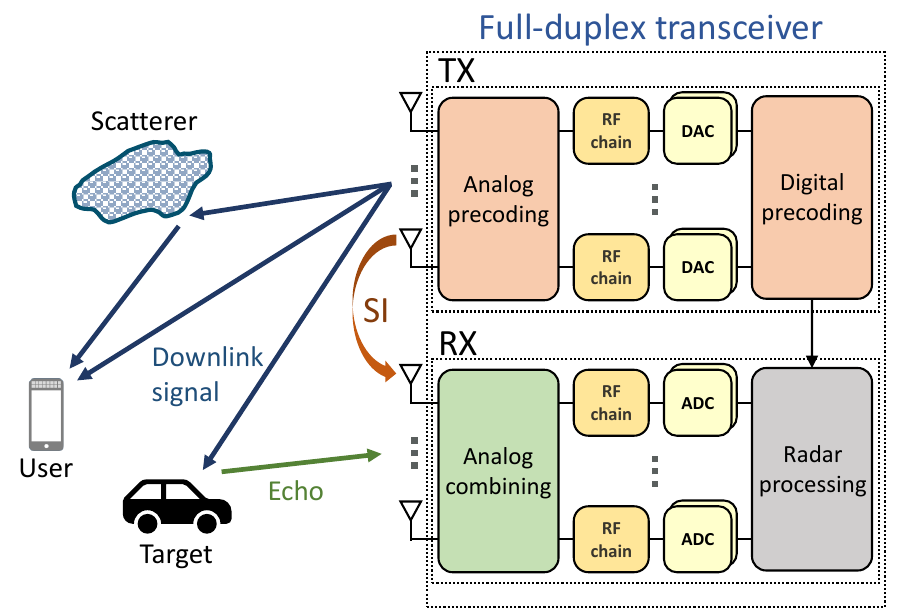}
    \vspace{-6pt}
    \caption{FD-JRC system with a downlink MS and target echoes.}
    \vspace{-12pt}
    \label{fig:diagram}    
\end{figure}

We consider the downlink of an FD-JRC system where the FD BS transmits $\Ns$ data streams to the MS, while the receiver side of the FD BS concurrently observes the reflections from targets. 
The main assumption is that the downlink channel and the target angles have been pre-estimated.
That is, a tracking scenario is considered for the radar operation such that the range and velocity of the targets have to be estimated, and the precoder and combiner have to be optimized for sensing and communication.
The illustration of the considered FD-JRC system is given in Fig~\ref{fig:diagram}.
Although there are multiple targets in the environment, we estimate each target in separate data frames. Since the frame duration is very short at mmWave, this does not introduce any practical limitation for sensing.

We assume OFDM modulation with $M$ subcarriers for signal transmission. 
To enable the FD operation for sensing, the BS has two collocated uniform linear arrays (ULAs) with $\NBS$ antennas.
The FD BS operates with a hybrid architecture with $\NBSRF$ radio frequency (RF) chains at both transmit and receive sides.
Furthermore, the MS employs a hybrid architecture with $\NMS$ antennas and $\NMSRF$ RF chains. 
The received signal at the MS, $\yMS \in \bbC^{\NMS}$, can be written as 
\begin{equation}\label{eq:yMS}
    \yMS = \left[\WMS\right]^\rmH \HBM \FBS \bs_{m,n} + \left[\WMS\right]^\rmH \nMS,
\end{equation}
for $m=0,\dots,M-1$ and $n=0,\dots,N-1$, where the number of OFDM symbols is denoted by $N$.
The transmit signal and the noise vector at the $m$-th subcarrier of the $n$-th OFDM symbol are denoted by $\bs_{m,n} \in \bbC^{\Ns}$, with covariance $\Exp{\bs_{m,n} \bs_{m,n}^\rmH} = \frac{\Pt}{\Ns}\bI_{\Ns}$, and $\nMS \in \bbC^{\NMS}$, with covariance $\Exp{\nMS \left[\nMS\right]^\rmH} = \varMS\bI_{\NMS}$, where $\Pt$ is the transmit power and $\varMS$ is the noise variance at the MS. 
Furthermore, the overall precoder at the BS for the $m$-th subcarrier is represented by $\FBS = \FBSRF \FBSBB \in \bbC^{\NBS \times \Ns}$, where $\FBSRF \in \bbC^{\NBS \times \NBSRF}$ is the frequency-flat analog precoder and $\FBSBB \in \bbC^{\NBSRF \times \Ns}$ is the digital precoder. 
Similarly, the overall combiner at the MS for the $m$-th subcarrier is represented by $\WMS = \WMSRF \WMSBB \in \bbC^{\NMS \times \Ns}$, where $\WMSRF \in \bbC^{\NMS \times \NMSRF}$ is the frequency-flat analog combiner and $\WMSBB \in \bbC^{\NMSRF \times \Ns}$ is the digital combiner.
The downlink channel at the $m$-th subcarrier is denoted by $\HBM \in \bbC^{\NMS \times \NBS}$. We assume that time and frequency synchronization is handled and the channel remains constant for $N$ OFDM symbols. Leveraging the geometric channel model \cite{Heath2016JSTSP}, and neglecting the impact of pulse shaping and filtering, the downlink channel can be expressed as
\vspace{-3pt}\begin{equation}\label{eq:HBM}
    \HBM = \sum_{l=1}^{L} \alphaBMl e^{-j2\pi m\tau_{l} \Delta f} \aMS(\phiBMl) \aBS^\rmH(\thetaBMl),\vspace{-3pt}
\end{equation}
where $L$ is the number of paths and $\Delta f$ is the subcarrier spacing. 
The complex gain, delay, angle-of-arrival (AoA) and angle-of-departure (AoD) of the $l$-th path are denoted by $\alphaBMl$, $\tau_{l}$, $\phiBMl$ and $\thetaBMl$, respectively. 
Furthermore, $\aMS(\phi) \in \bbC^{\NMS}$ is the array response at the MS for the incident angle $\phi$ and $\aBS(\theta) \in \bbC^{\NBS}$ is the array response at the BS for the incident angle $\theta$.
On the other hand, the received signal at the BS, $\yBS \in \bbC^{\NBSRF}$ can be expressed as
\vspace{-3pt}\begin{multline}
    \yBS = \left[\WBS\right]^\rmH \Ht \FBS \bs_{m,n} \\
    + \sqrt{\rho} \left[\WBS\right]^\rmH \HSI \FBS \bs_{m,n} + \left[\WBS\right]^\rmH \nBS,
\end{multline}
where $\rho$ is the SI channel power, $\WBS \in \bbC^{\NBS \times \NBSRF}$ is the analog combiner at the BS, $\Ht \in \bbC^{\NBS \times \NBS}$ is the channel associated with the targets at the $m$-th subcarrier of the $n$-th OFDM symbol, $\HSI \in \bbC^{\NBS \times \NBS}$ is the SI channel and $\nBS \in \bbC^{\NBS}$ is the noise vector at the BS with covariance $ \Exp{\nBS \left[\nBS\right]^\rmH} = \varBS\bI_{\NBS} $ and noise variance $\varBS$. 
We consider $K$ point targets in the environment. 
Since the transmit and receive arrays of the FD BS are collocated, the AoA and AoD of each target are assumed to be the same \cite{Barneto2022TCOM, Islam2022ICC, Bayraktar2023CAMSAP}. Therefore, the target channel matrix can be expressed as
\vspace{-3pt}\begin{equation}
    \Ht = \sum_{k=1}^{K}\alphat e^{j2\pi (n T f_{D,k} - m\tau_{k}^\mathrm{t} \Delta f) } \aBS(\thetat) \aBS^\rmH(\thetat),
\end{equation}
where $\alphat$ is the complex reflection coefficient, $f_{D,k}$ is the Doppler frequency, $\tau_{k}^\mathrm{t}$ is the round-trip time of the reflection and $\thetat$ is the AoA/AoD of the $k$-th target. 
In addition, $T$ is the duration of an OFDM symbol.
Finally, the SI channel between the transmit and receive arrays of the BS is modeled with the near-field LoS channel model as \cite{Islam2022ICC, Bayraktar2023CAMSAP}
\begin{equation}\label{eq:HSI}
    \left[\HSI\right]_{p,q} = \frac{\gamma}{d_{pq}} e^{-j2\pi\frac{d_{pq}}{\lambda}}, \quad p,q=1,\dots,\NBS,
\end{equation}
where $d_{pq}$ is the distance between the $p$-th receive antenna and the $q$-th transmit antenna, $\lambda$ is the wavelength and $\gamma$ is the normalization constant to satisfy $ \norm{\HSI}_\rmF^2 = \NBS^2 $.

\section{Precoder/combiner Design for FD-JRC}
\label{sec:design}
FD-JRC systems require maintaining a trade-off between communication and sensing performance while mitigating the SI for radar operation.
We achieve this goal by designing the precoder and combiner at the FD BS.
We also design a protocol such that one of the targets is estimated at each OFDM frame.
In this section, we first introduce the metrics for communication and sensing performance. 
Then, we propose an algorithm that maximizes the communication performance under the sensing and SI constraints.

The communication performance metric is the spectral efficiency for the $m$-th subcarrier, written as
\begin{equation}
    \cR_{m} \!\! = \! \log_{2} \left| \bI_{\Ns} \!\! + \! \frac{\Pt}{\varMS \Ns} \! \left[\WMS\right]^{\dagger} \! \HBM \FBS \! \left[\FBS\right]^{\rmH} \! \HBM^{\rmH} \WMS \right|,
\end{equation}
where $\left[\cdot\right]^{\dagger}$ is the Moore-Penrose pseudoinverse.
The radar gain achieved with a transmit precoder $\FBS$ is defined as
\begin{equation}
	\GT(\thetar) = \left| \aBS^\rmH(\thetar) \left[\FBS\right]_{:,\ns} \right|,
\end{equation}
for $\ns=1,\dots,\Ns$. Similarly, the radar gain at the receiver side of the FD BS can be defined as 
\begin{equation}
\GR(\thetar) = \left|\left[\WBS\right]_{:,\nRF}^\rmH \aBS(\thetar)\right|,
\end{equation}
for $\nRF=1,\dots,\NBSRF$. Finally, the radar signal-to-interference-plus-noise ratio (SINR) observed at the $\nRF$-th RF chain of the receiver side of the BS is defined as
\begin{equation}
    \SINR \! \! = \! \! \frac{\frac{\Pt}{\Ns} \norm{\left[\WBS\right]_{:,\nRF}^\rmH \Ht \FBS }_{\rmF}^2 }{\frac{\Pt\rho}{\Ns} \! \norm{\left[\WBS\right]_{:,\nRF}^\rmH \! \! \HSI \FBS}_{\rmF}^2 \! + \! \varBS \! \norm{\left[\WBS\right]_{:,\nRF} \!}^2}.
\end{equation}
Our aim is to design the precoders and combiners such that the downlink spectral efficiency is maximized under the condition that transmit and receive radar gains are high, while the SI is mitigated by precoders/combiners that consider the unit magnitude constraint for the entries of their analog stages. This approach effectively provides high radar SINR which is closely related to the quality of the target estimation \cite{Liu2022JSAC}. Following this description, the optimization problem for the precoders and combiners can be stated as follows:
\begin{equation}\label{eq:problem_overall}
    \begin{aligned}
    & \underset{ \substack{\FBS, \WBS, \\ \WMS}}{\text{maximize}} & & \sum_{m=0}^{M-1} \cR_{m}\\
    & \text{subject to} & &  \GT(\thetar) \geq \tauT, \: \forall m, \ns, \\
    & & & \GR(\thetar) \geq \tauR, \: \forall \nRF, \\
    & & & \left[\WBS\right]^\rmH  \HSI \FBS = \boldsymbol{0}, \: \forall m, \\
    & & & \left|\left[\WBS\right]_{i,j}\right| = 1, \: \forall i,j, \\
    & & & \norm{\FBS}_\rmF^{2} = \Ns, \: \forall m,
\end{aligned}
\end{equation}
where the transmit power is fixed with the last constraint. Furthermore, $\tauT$ and $\tauR$ are the transmit and receive radar gain thresholds.
Note that the unit modulus constraints for the analog stage of the hybrid precoder at the BS and the hybrid combiner at the MS are not imposed at this stage. 
Even without these constraints, the problem is non-convex and hard to solve, mainly due to the unit modulus constraint of the combiner at the BS. 
Thus, we employ an alternating optimization approach to find each variable while fixing others.
In the rest of the section, we introduce and solve each subproblem to obtain  the fully digital solutions, which are later decomposed into their corresponding hybrid approximations. 

\subsection{Precoder Optimization for the BS}
The precoder at the BS has to maximize the spectral efficiency across all subcarriers while mitigating the SI and having a high radar gain in the target direction.
For fixed combiners, the problem in \eqref{eq:problem_overall} can be reduced to
\begin{equation}\label{eq:problem_precoder}
    \begin{aligned}
    & \underset{\FBS}{\text{maximize}} & & \sum_{m=0}^{M-1} \cR_{m}\\
    & \text{subject to} & & \GT(\thetar) \geq \tauT, \: \forall m,\ns, \\
    & & & \left[\WBS\right]^\rmH  \HSI \FBS = \boldsymbol{0}, \: \forall m, \\    
    & & & \norm{\FBS}_\rmF^{2} = \Ns, \: \forall m.
\end{aligned}
\end{equation}
The hybrid analog/digital constraint of the precoders is initially omitted, however, will be satisfied after finding the fully digital precoders.
There are important observations related to the optimization problem in \eqref{eq:problem_precoder}.
Firstly, optimization of fully digital precoders for different subcarriers are independent of each other with fixed analog combiners.
Furthermore, the power constraint can be omitted for now, and can be enforced later by normalization. In this case, the optimization problem for the precoder at the $m$-th subcarrier can be written as
\begin{equation}\label{eq:problem_precoder_subcarrier}
    \begin{aligned}
    & \underset{\FBS}{\text{maximize}} & & \cR_{m}\\
    & \text{subject to} & & \GT(\thetar) \geq \tauT, \: \forall \ns, \\
    & & & \left[\WBS\right]^\rmH  \HSI \FBS = \boldsymbol{0}.
\end{aligned}
\end{equation}
It is well known that the maximum spectral efficiency is achieved by setting the precoder as the $\Ns$ right singular vectors of $\HBM$ that correspond the the largest singular values, if the optimal combiner is used at the MS \cite{Heath2016JSTSP}. This is equivalently achieved by selecting the first $\Ns$ eigenvectors of $\bA_{m} = \HBM^\rmH \HBM$, which corresponds to the maximization of $\trace{\bF_{m}^\rmH \bA_{m} \bF_{m}}$ with the constraint $\bF_{m}^\rmH \bF_{m} = \bI_{\Ns}$, where $\bF_{m} \in \bbC^{\NBS \times \Ns}$ is the optimization variable. Let us omit the transmit gain constraint in \eqref{eq:problem_precoder_subcarrier} for now. In this case, we consider the maximization of the spectral efficiency while minimizing the SI term. One way to achieve this goal is to fix the SI term and maximize the spectral efficiency, which can be expressed as
\begin{equation}\label{eq:generalized_eigenvalue_Fm}
	\begin{aligned}
		& \underset{ \substack{\bF_{m}}}{\text{maximize}} & & \trace{\bF_{m}^\rmH \bA_{m} \bF_{m}}\\
		& \text{subject to} & & \bF_{m}^\rmH \bC \bF_{m} = \bI,
	\end{aligned}
\end{equation}
where $ \bC = \HSI^\rmH \WBS \left[\WBS\right]^\rmH \HSI $ is the covariance matrix of the SI space that $\bF_{m}$ should mitigate.
The problem in \eqref{eq:generalized_eigenvalue_Fm} is shown to be equivalent to $ \bA_{m} \bF_{m} = \bC \bF_{m} \boldsymbol{\Lambda}_{m}
$ where the columns of $\bF_{m}$ are the eigenvectors of $\bA_{m}$, and the diagonal entries of $\boldsymbol{\Lambda}_{m}$ are the corresponding eigenvalues \cite{Ghojogh2019}. 
This expression is called the generalized eigenvalue problem.
We utilize the solution to the generalized eigenvalue problem to maximize the spectral efficiency while mitigating the SI.
It is worth noting that generalized eigenvalue problem-based beamforming has found applications in various communication scenarios, such as signal-to-leakage ratio-based multiuser precoding \cite{Sadek2007}, cooperative multicell beamforming \cite{Bhagavatula2010}, and spatial user grouping-based beamforming \cite{Bayraktar2021}.

To satisfy the target gain constraint, we consider the coherent combination of $\bF_{m}$ with the beams that have high gain in the target direction \cite{Barneto2022TCOM, Elbir2023SPM}. 
Note that these beams should also minimize the SI so that the coherent combination results in low residual SI. To that end, we can construct another generalized eigenvalue problem expressed as
\begin{equation}\label{eq:generalized_eigenvalue_f}
	\begin{aligned}
		& \underset{ \substack{\bff}}{\text{maximize}} & & \trace{\bff^\rmH \bB \bff}\\
		& \text{subject to} & & \bff^\rmH \bC \bff = 1,
	\end{aligned}
\end{equation}
where $ \bB = \aBS(\thetar) \aBS^\rmH(\thetar) $ is the spatial covariance matrix of the target direction.
Since the solutions to both problems suppress SI, we can coherently combine them as
\begin{equation}\label{eq:coherent}
    \left[\FBS\right]_{:,n_{\rm s}} = \kappa \left[\bF_{m}\right]_{:,n_{\rm s}} + (1-\kappa) \bff,
\end{equation}
for $n_{\rm s}=1,\dots,\Ns$. Then, we can normalize the precoder to satisfy $\norm{\FBS}_{\rm F}^2 = \Ns$. 
The compromise between communication and sensing is satisfied by $\kappa$ such that $\GT(\thetar) = \tauT$, where $0\leq\kappa\leq 1$. 
Note that the described steps provide a fully digital solution to the problem \eqref{eq:problem_precoder}.
We resort to the hybrid decomposition of the fully digital precoders \cite{Elbir2023SPM}. Considering that the analog precoder is frequency-flat, the hybrid decomposition across all subcarriers is given as
\begin{equation}\label{eq:hybrid}
	\begin{aligned}
		& \underset{\mathclap{\FBSRF,\FBSBB}}{\text{minimize}} \quad & & \sum_{m=0}^{M-1} \norm{\FBS-\FBSRF\FBSBB}_\rmF^{2}\\
		& \text{subject to} & & \left| \left[\FBSRF\right]_{i,j} \right| = 1, \: \forall i,j,\\
        & & & \norm{\FBSRF\FBSBB}_\rmF^{2} = \Ns, \: \forall m.
	\end{aligned}
\end{equation}
There are many algorithms developed in the literature to solve this problem \cite{Heath2016JSTSP,Rusu2016TWC,Yu2016JSTSP,Coma2018JSTSP}.
We use the low-complexity phase extraction with alternating minimization method in \cite{Rusu2016TWC,Yu2016JSTSP} to solve the problem \eqref{eq:hybrid}. 
While the proposed fully digital solution provides perfect SI suppression, the hybrid decomposition would degrade the suppression capability. Thus, we further suppress the SI with the combiner at the BS.

\subsection{Combiner Optimization for the MS}
The combiners at the MS, $\WMS$, are optimally selected as the $\Ns$ right singular vectors of the channels $\HBM$. 
We use the same hybrid decomposition algorithm to find the analog and digital combiners, as done for the precoders in \eqref{eq:hybrid}.

\subsection{Combiner Optimization for the BS}
The combiner optimization aims to minimize the total SI over all the subcarriers while maintaining the gain in the target direction above a threshold. For given precoders, the optimization problem can be explicitly stated as
\begin{equation}\label{eq:combiner_problem}
    \begin{aligned}
    & \underset{\WBS}{\text{minimize}} & & \sum_{m=0}^{M-1} \norm{\left[\WBS\right]^\rmH \HSI \FBS}_{\rmF}^{2}\\
    & \text{subject to} & & \GR(\thetar) \geq \tauR, \: \forall \nRF, \\
    & & & \left|\left[\WBS\right]_{i,j}\right| = 1, \: \forall i,j,
\end{aligned}
\end{equation}
where the unit-modulus constraint makes the problem non-convex. Firstly, we need to modify the first constraint as it is a concave function. Instead of forcing the radar gain to be above a threshold, we limit the loss of gain which is convex. Then, we relax the unit-modulus constraint by letting the entries of the combiner $\WBS$ deviate from the unit circle in a controllable manner. We use a block coordinate descent approach to solve the problem for a certain number of the entries which are randomly selected at each iteration as done in \cite{Bayraktar2023CAMSAP}. The optimization process is terminated when the cost function converges. Furthermore, we limit the deviation of the variables from the previous iteration to avoid abrupt changes. The convex problem solved at the $n$-th iteration is given as
\begin{equation}\label{eq:problem2}
    \begin{aligned}
    & \underset{\left[\WBS\right]_{\cI_n}}{\text{minimize}} & & \sum_{m=0}^{M-1} \norm{\left[\WBS\right]^\rmH \HSI \FBS}_{\rmF}^{2}\\
    & \text{subject to} & & \left|\NBS - \left[\WBS\right]_{:,\nRF}^\rmH \aBS(\thetar) ) \right| \leq (\NBS - \tauR), \\ 
    & & & \left|\left[\WBS\right]_{i,j}\right| - 1 \leq \epsilon_1, \: (i,j) \in \cI_n, \\
    & & & \norm{\left[\WBS\right]_{\cI_n} - \left[\WBS_{(n-1)}\right]_{\cI_n}  } \leq \epsilon_2,
\end{aligned}
\end{equation}
where $\epsilon_1$ and $\epsilon_2$ are small threshold values. In addition, $\cI_n$ is the set of randomly selected indices from $\WBS$ to be optimized at the $n$-th iteration. The optimized entries at the $n$-th iteration are denoted by $ \left[\WBS\right]_{\cI_n} \in \bbC^{|\cI_n|} $, while the combiner found in the previous iteration is represented by $\WBS_{(n-1)}$. The entries are normalized to unity after each iteration is finalized to satisfy the unit-modulus constraint. Furthermore, the columns of the combiner are initialized with $\aBS(\thetar)$. The problem \eqref{eq:problem2} is solved with the convex solver CVX \cite{Grant2014cvx}. The provided alternating optimization solution ensures convergence to local minima as shown in \cite{Bayraktar2021, Elbir2023SPM, Rusu2016TWC, Yu2016JSTSP, Coma2018JSTSP}.

%
\section{Numerical Results}
\label{sec:results}
In this section, we evaluate the proposed precoder/combiner design for FD-JRC at mmWave. 
Note that the results are shown without considering the saturation and clipping effects at the receiver of the BS in order to observe the degradation in SI suppression.
Regarding the system parameters, the BS is equipped with two collocated ULAs with $\NBS = 32$ antennas parallel to the x-axis.
The separation between the receive and transmit arrays is $ 6\lambda $ in the z-axis.
The number of RF chains at the BS is $\NBSRF=4$.
The MS has an ULA with $\NMS=16$ antennas and $\NMSRF=4$ RF chains. 
The number of streams is set to $\Ns=4$.
The spacing between the antenna elements for all the arrays is half wavelength.
The system operates at $28$GHz with $100$MHz bandwidth that corresponds to a noise power of $-93.8$dBm at room temperature. 
The number of active subcarriers and OFDM symbols are set to $M=792$ and $N=14$, following the 5G NR standards with subcarrier spacing $\Delta f = 120$kHz and symbol duration $T = 8.92\mu$s \cite{Islam2022ICC}. In the combiner design problem for the BS, the receive gain threshold is set to $\tauR = 0.7\NBS$, and the design parameters are set to $\epsilon_1=0.3$ and $\epsilon_2=0.1$.

\subsection{Spectral Efficiency and Radar SINR Evaluation}
We evaluate first the spectral efficiency performance of the designed precoders. We consider a point target with radar cross-section of $10\textrm{m}^2$ at azimuth angle $\thetar = 45^\circ$ and a distance of $40$m from the BS. The MS is randomly deployed at a distance of $50$m from the BS with the LoS angle selected from $[-60^\circ,60^\circ]$. The downlink channel has $L=5$ paths. The gains of non-LoS paths are $5$-$15$dB below the gain of the LoS path, while the angles are randomly selected from $[-90^\circ,90^\circ]$. The results are averaged over $100$ different realizations. We use two benchmarks which do not account for SI. The first one is the optimal precoding that uses the right singular vectors of $\HBM$ providing the maximum spectral efficiency. The second one is the coherent eigenvector method that uses the combination of the eigenvectors from the problems \eqref{eq:generalized_eigenvalue_Fm} and \eqref{eq:generalized_eigenvalue_f} without the SI constraint, while utilizing the same hybrid decomposition algorithm.

\begin{figure}
\centering
\includegraphics[width=0.98\linewidth]{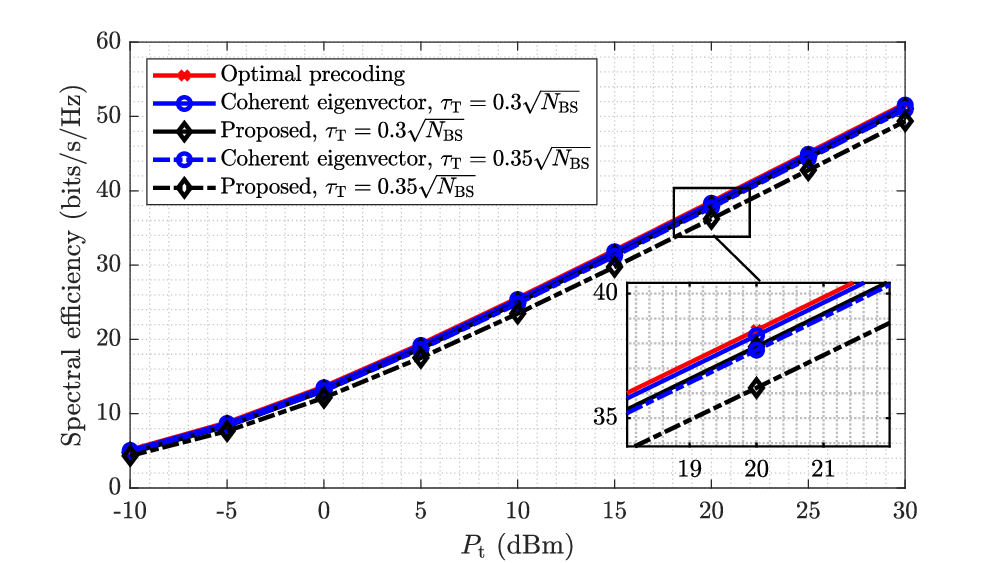}
\vspace{-8pt}
\caption{Downlink spectral efficiency vs $\Pt$ for different precoding methods.}
\vspace{-13pt}
\label{fig:spectral}
\end{figure}

\begin{figure}
\centering
\includegraphics[width=0.98\linewidth]{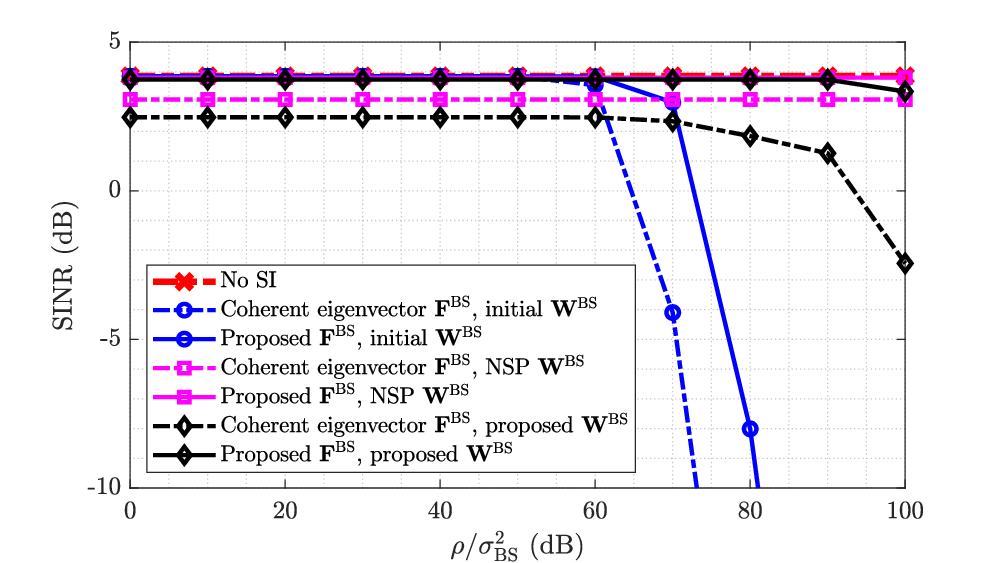}
\vspace{-8pt}
\caption{Radar SINR vs SI-to-noise ratio for different precoding and combining methods with $\tauT=0.35\sqrt{\NBS}$ and $\Pt=20$dBm.}
\vspace{-10pt}
\label{fig:SINR}
\end{figure}

The spectral efficiency curves with respect to the transmit power for the two benchmark schemes and the proposed algorithm with $\tauT=0.3\sqrt{\NBS}$ and $\tauT=0.35\sqrt{\NBS}$ are given in Fig.~\ref{fig:spectral}.
It can be seen that the spectral efficiency loss is negligible in the case of $\tauT=0.3\sqrt{\NBS}$ for both coherent eigenvector and the proposed method. 
If the gain threshold is increased to $\tauT=0.35\sqrt{\NBS}$, we observe slight degradation for the coherent eigenvector method since it allocates more power to the target direction. 
We also observe more degradation for the proposed method as it also suppresses the SI while maintaining the required transmit gain. 
In turn, the proposed precoder provides higher radar SINR, especially at high SI-to-noise levels. 
To demonstrate this fact, we show the radar SINR with respect to SI-to-noise ratio with $\tauT=0.35\sqrt{\NBS}$ and $\Pt=20$dBm in Fig.~\ref{fig:SINR}.
The no SI case is obtained assuming there is no SI while using the coherent eigenvalue precoder and the initial $\WBS$. 
It can be seen that the proposed precoder provides high SINR even at high SI-to-noise levels, while the coherent eigenvector method cannot provide the required SINR in any case. 
Furthermore, we also design the combiner at the BS by using the NSP approach proposed in \cite{Barneto2022TCOM} as another benchmark. 
This approach results in a fully digital combiner  whose approximate implementation requires of a complicated analog hardware. 
While the NSP combiner completely mitigates the SI, it loses from the target gain when coherent eigenvector is used for the precoder.
Moreover, the proposed combiner implemented with only phase shifters attains the same performance as NSP even at high SI-to-noise ratio levels when the proposed precoder is utilized.

\subsection{Range and Velocity Estimation for the Targets}
We consider the subcarrier-based OFDM radar processing for range and velocity estimation \cite{Barneto2022TCOM, Islam2022ICC}. 
To that end, we construct the matrix $\bZ \in \bbC^{M \times N}$ such that the entries are 
\begin{equation}
\left[\bZ\right]_{m,n} = 
\sum_{\nRF=1}^{\NBSRF} \frac{\left[\yBS\right]_{\nRF}  }{\left(\left[\WBS\right]_{:,\nRF}^\rmH   \aBS(\thetar) \aBS^\rmH(\thetar) \FBS \bs_{m,n}  \right)^{*}}.
\end{equation}
The subcarriers have a progressive phase shift caused by the round-trip time and the OFDM symbols have a progressive phase shift due to the Doppler effect. Therefore, the DFT of $\bZ$ across OFDM symbols, and then the inverse DFT across subcarriers are taken to obtain the sparse image $\bar{\bZ} \in \bbC^{\bar{M} \times \bar{N}}$, where $\bar{M}$ and $\bar{N}$ are the sizes of the DFT and inverse DFT operations, respectively. The maximum entry of $\bar{\bZ}$ contains the information related to range and velocity, i.e., $(\bar{m}^{\star}, \bar{n}^{\star}) = \argmax_{\bar{m},\bar{n}} \big|\left[\bar{\bZ}\right]_{\bar{m},\bar{n}}\big|$ for $\bar{m}=0,\dots,\bar{M}-1$ and $\bar{n}=0,\dots,\bar{N}-1$. The range and the velocity of the target can be recovered as $r=\frac{\bar{m}^{\star}c}{2\bar{M}\Delta f}$ and $v=\frac{\bar{n}^{\star}\lambda}{\bar{N}T}$, respectively.

\begin{figure}
\centering
\subfloat[Angle-range map]{\includegraphics[width=0.85\linewidth]{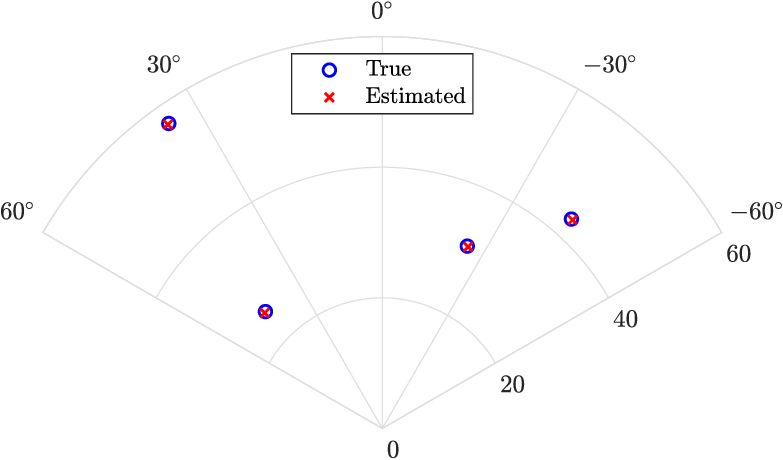}}
\vspace{-1pt}
\subfloat[Range-velocity map]{\includegraphics[width=0.9\linewidth]{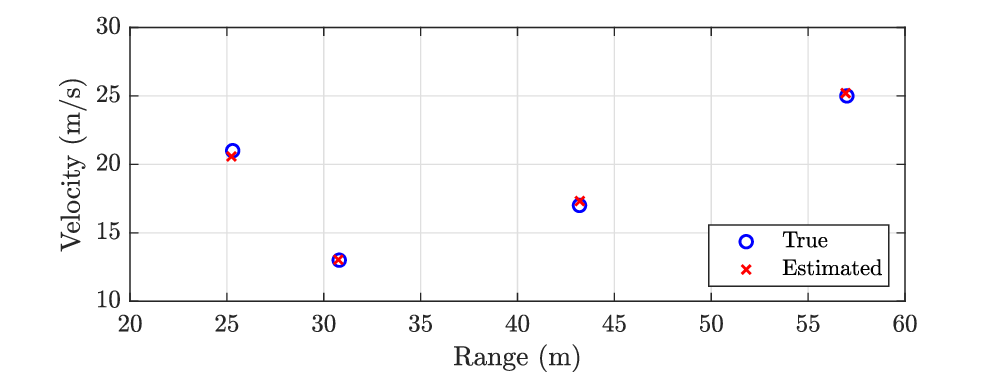}}
\caption{Angle-range and range-velocity maps of the targets.}
\vspace{-15pt}
\label{fig:sensing}
\end{figure}

To demonstrate the performance of the proposed algorithm, we construct the precoders and combiner at the BS for each target angle. Then, the described range and velocity estimation procedure is applied. We use the same setting as in previous subsection with transmit gain threshold $\tauT=0.35\sqrt{\NBS}$, transmit power $\Pt=20$dBm and SI-to-noise ratio $\rho/\varBS=80$dB. There is an MS with LoS angle $10^{\circ}$ at a distance of $50$m from the BS. There are $4$ point targets with radar cross-section of $10\textrm{m}^2$. The inverse DFT and DFT sizes are set to $\bar{M} = 10M$ and $\bar{N} = 500N$ for the range and velocity estimation, respectively. The obtained angle-range and range-velocity maps are given in Fig.~\ref{fig:sensing}. It is observed that the range and velocity estimates are highly accurate, which is an indication that the SI is efficiently mitigated and the radar gain is sufficient. Note that the angle estimates are assumed to be previously obtained with error lower than $0.5^{\circ}$.


\section{Conclusion}
\label{sec:conc}
In this paper, we proposed a precoder/combiner design for FD-JRC at mmWave, utilizing a generalized eigenvalue-based precoding to suppress SI, maximize downlink spectral efficiency and guarantee certain SINR for radar.
To further suppress the SI, we introduced a non-convex design problem for the analog combiner, which is solved with block coordinate descent and convex relaxation techniques. 
Simulation results demonstrate the effectiveness of our approach against SI. 
Additionally, we employ OFDM-based radar processing for range and velocity estimation, showcasing high accuracy with the proposed architecture.

\bibliographystyle{IEEEtran}
\bibliography{refs}

\end{document}